# SPH CALCULATIONS OF COLLISIONS BETWEEN MAIN-SEQUENCE STARS

FREDERIC A. RASIO

*Department of Physics, M.I.T. 6-201, Cambridge, MA 02139, USA*

**Abstract.** The hydrodynamics of collisions and mergers of main-sequence stars is discussed in the light of recent 3-D calculations using the smoothed particle hydrodynamics (SPH) method. Theoretical models for the formation of blue stragglers are reviewed in the context of recent comparisons between the observed properties of blue stragglers in dense globular clusters and the predictions of those models.

## 1. Introduction

Close dissipative encounters and direct physical collisions between stars occur frequently in dense star clusters. The dissipation of kinetic energy in close stellar encounters can have a direct influence on the dynamical evolution of a cluster, since it encourages secular core collapse. At the same time, however, mass loss due to accelerated stellar evolution in merger products tends to *unbind* the parent system (Spitzer 1987; Statler et al. 1987; Goodman & Hernquist 1991). Observational evidence for stellar collisions and mergers in globular clusters is provided by the existence of large numbers of blue stragglers in these systems. Blue stragglers are main-sequence stars that appear above the turnoff point in the color-magnitude diagram of a cluster. They have long been thought to be formed through the merger of two lower-mass stars, either in a collision or following binary coalescence (Leonard 1989; Livio 1993; Stryker 1993; Bailyn 1995). Clear indication for a collisional origin of blue stragglers in dense globular clusters has come from recent observations of cluster cores by the Hubble Space Telescope. Large numbers of blue stragglers were found to be concentrated in the cores of the densest clusters, such as M15 (De Marchi & Paresce 1994; Guhathakurta et al. 1995) and M30 (Yanny et al. 1994).

Collisions can happen directly between two single stars only in the cores of the densest clusters, but even in somewhat lower-density clusters they can also happen indirectly, during resonant interactions involving wide primordial binaries (Leonard 1989; Sigurdsson & Phinney 1995; Davies & Benz 1995). The existence of dynamically significant numbers of primordial binaries in globular clusters is now well established observationally (Hut et al. 1992; Cote et al. 1994; see also the article by Pryor in this volume). In dense cluster cores, close binaries, perhaps formed by tidal capture, may be quickly destroyed by interactions with other stars or binaries, also leading to collisions (Goodman & Hernquist 1991; see also the articles by Mardling and by Shara in this volume).

Following early numerical work in 2-D (e.g., Shara & Shaviv 1978), Benz & Hills (1987, 1992) performed the first 3-D calculations of direct collisions between two main-sequence stars. An important conclusion of their pioneering study was that stellar collisions could lead to thorough mixing of the fluid. In particular, they pointed out that the mixing of fresh hydrogen fuel into the core of the merger remnant could reset the nuclear clock of a blue straggler, allowing it to remain visible for a full main-sequence lifetime $t_{MS} \sim 10^9$ yr after its formation.

In subsequent work it was generally assumed that the merger remnants resulting from stellar collisions were nearly homogeneous. Blue stragglers would then start their life close to the zero-age main sequence, but with an anomalously high helium abundance coming from the hydrogen burning in the parent stars. In contrast, little hydrodynamic mixing was expected to occur during the much gentler process of binary coalescence, which could take place on a stellar evolution timescale rather than on a dynamical timescale (Mateo et al. 1990; Bailyn 1992; but see Rasio 1993, 1995, and Rasio & Shapiro 1995).

On the basis of these ideas, Bailyn (1992) suggested a way of distinguishing observationally between the two possible formation processes. The helium abundance in the envelope of a blue straggler, which reflects the degree of mixing during its formation process, can affect its observed position in a color-magnitude diagram. Blue stragglers made from collisions would have a higher helium abundance in their outer layers than those made from binary mergers, and this would generally make them appear somewhat brighter and bluer.

A detailed analysis was carried out by Bailyn & Pinsonneault (1995) who performed stellar evolution calculations for blue stragglers assuming various initial chemical composition profiles. To represent the collisional case, they assumed chemically homogeneous initial profiles with enhanced helium abundances, calculating the total helium mass from stellar evolution models of the parent stars. For the dense cluster 47 Tuc they concluded that the observed luminosity function and numbers of blue stragglers were then consistent with a collisional origin.

## 2. The SPH Method

The vast majority of recent 3-D calculations of stellar interactions (collisions, binary coalescence, common envelope evolution, tidal disruption, etc.) have been done using the smoothed particle hydrodynamics (SPH) method (see Monaghan 1992 for a recent review). Since SPH is a Lagrangian method, in which particles are used to represent fluid elements, it is ideally suited for the study of hydrodynamic mixing. Indeed, chemical abundances are passively advected quantities during the dynamical evolution. Therefore, the chemical composition in the final fluid configuration can be determined after the completion of a calculation simply by noting the original and final positions of all SPH particles and by assigning particle abundances according to an initial profile.

A straightforward derivation of the basic SPH equations can be obtained from a Lagrangian formulation of hydrodynamics (Gingold & Monaghan 1982). Consider for simplicity an ideal fluid undergoing adiabatic evolution. The Euler equations of motion,

$$\frac{d\mathbf{v}}{dt} = \frac{\partial \mathbf{v}}{\partial t} + (\mathbf{v} \cdot \nabla)\mathbf{v} = -\frac{1}{\rho}\nabla p, \qquad p = A\rho^\gamma, \qquad (1)$$

can be derived from a Lagrangian principle with

$$L = \int \left\{ \frac{1}{2}\dot{\mathbf{x}}^2 - u[\rho(\mathbf{x})] \right\} \rho \, d^3x. \qquad (2)$$

Here $p$ is the pressure, $\rho$ is the density, $A \propto \exp(s)$ is a function of the specific entropy $s$ (assumed here to be constant in space and time), and $u[\rho] = p/[(\gamma-1)\rho] = A\rho^{\gamma-1}/(\gamma-1)$ is the specific internal energy of the fluid.

The basic idea in SPH is to use the discrete representation

$$L_{SPH} = \sum_{i=1}^{N} m_i \left[ \frac{1}{2}\dot{\mathbf{x}}_i^2 - u(\rho_i) \right] \qquad (3)$$

for the Lagrangian, where the sum is over a large but discrete number of small cells, or "particles," covering the volume of the fluid. Here $m_i$ is the mass of a particle, $\mathbf{x}_i$ is its position, and $\dot{\mathbf{x}}_i$ is its velocity. For expression (3) to become the Lagrangian of a system with a finite number $N$ of degrees of freedom, we need a prescription to compute the density $\rho_i = \rho(\mathbf{x}_i)$ at the position of a given particle, as a function of the masses and positions of neighboring particles. In SPH, this is done by introducing a local average,

$$\rho_i = \sum_j m_j W_{ij}, \qquad W_{ij} = W(|\mathbf{x}_i - \mathbf{x}_j|; h), \qquad (4)$$

where $W(r; h)$ is a smoothing kernel, normalized to unity and of width $\sim h$. A very common choice for the smoothing kernel is the cubic spline

$$W(r; h) = \frac{1}{\pi h^3} \begin{cases} 1 - \frac{3}{2}\left(\frac{r}{h}\right)^2 + \frac{3}{4}\left(\frac{r}{h}\right)^3, & 0 \leq \frac{r}{h} < 1, \\ \frac{1}{4}\left[2 - \left(\frac{r}{h}\right)\right]^3, & 1 \leq \frac{r}{h} < 2, \\ 0, & \frac{r}{h} \geq 2. \end{cases} \qquad (5)$$

(Monaghan & Lattanzio 1985). With the prescription (4) for the density, we can now obtain the equations of motion for all the particles. Deriving the Euler-Lagrange equations from $L_{SPH}$ we get

$$\frac{d\mathbf{v}_i}{dt} = -\sum_j m_j \left(\frac{p_i}{\rho_i^2} + \frac{p_j}{\rho_j^2}\right) \nabla_i W_{ij}. \qquad (6)$$

The expression in the right-hand side of equation (6) is a sum over neighboring particles (within a distance $\sim h$ of $\mathbf{x}_i$) representing a discrete approximation to the pressure force $[-(1/\rho)\nabla p]_i$ acting on the particle at $\mathbf{x}_i$. Typically, a full implementation of SPH for astrophysical problems would add to equation (6) a treatment of self-gravity (e.g., using one of the many grid-based or tree-based algorithms developed for N-body simulations) and an artificial viscosity term to allow for entropy production in shocks. In addition, we have assumed here in deriving equation (6) that the smoothing length $h$ is constant in time and the same for all particles. In reality, individual and time-varying smoothing lengths $h_i(t)$ are almost always used, so that the local spatial resolution can be adapted to the (time-varying) density of SPH particles (see Nelson & Papaloizou 1994 for a rigorous derivation of the equations of motion in this case).

The following energy and momentum conservation laws are satisfied *exactly* by the simple SPH equations of motion given above

$$\frac{d}{dt}\left(\sum_{i=1}^N m_i \mathbf{v}_i\right) = 0, \qquad (7)$$

and

$$\frac{d}{dt}\left(\sum_{i=1}^N m_i [\frac{1}{2}v_i^2 + u_i]\right) = 0, \qquad (8)$$

where $u_i = p_i/[(\gamma - 1)\rho_i]$. Note that energy and momentum conservation in SPH is independent of the number of particles $N$.

## 3. Discussion of Recent Results

Lombardi, Rasio, & Shapiro (1995a,b) have studied collisions between main-sequence stars, and, in particular, the question of mixing during mergers, by performing a new set of numerical hydrodynamic calculations using SPH. This new work improves on the previous study of Benz & Hills (1987) by adopting more realistic stellar models, and by performing numerical calculations with increased spatial resolution. Benz & Hills (1987) used an early version of the SPH method and performed their calculations with a small number of particles ($N = 1024$). They also represented all stars by simple $n = 1.5$ polytropes. Unfortunately, $n = 1.5$ polytropes have density profiles that are not steep enough to represent main-sequence stars close to the turnoff point. Turnoff main-sequence stars have very shallow convective envelopes and are much better modeled by $n = 3$ polytropes (which have much more centrally concentrated density profiles).

The new SPH calculations of Lombardi et al. (1995a,b) are done using $N = 3 \times 10^4$ particles, and the colliding stars are modeled as composite polytropes (with $n = 3$ for the radiative interior and $n = 1.5$ in the convective envelope), which provide accurate representations of the density profiles in the entire mass range of interest for globular clusters (cf. Rappaport, Verbunt, & Joss 1983; Ruciński 1988). This is particularly important for collisions between two stars of different masses, which in general will also have different internal structures (and, for this reason, the later calculations of Benz & Hills 1992, done for two $n = 1.5$ polytropes with a mass ratio of 1/5, are of very limited applicability).

Stars close to the main-sequence turnoff point in a cluster are in fact the most relevant ones to consider for stellar collision calculations. Indeed, as the cluster evolves via two-body relaxation, the most massive stars will tend to concentrate in the dense cluster core, where the collision rate is highest (see, e.g., Spitzer 1987). In addition, collision rates can be increased dramatically by the presence of a significant fraction of primordial binaries in the cluster, and the more massive stars will preferentially tend to be exchanged into such a binary, or collide with another star, following a dynamical interaction between two binaries or between a binary and a single star (Sigurdsson & Phinney 1995).

The main new results of Lombardi et al. (1995a,b) can be summarized as follows. Typical merger remnants produced by collisions are far from chemically homogeneous. In the case of collisions between two nearly identical main-sequence stars close to the turnoff point, the amount of hydrodynamic mixing during the collision is minimal. In fact, the final chemical composition profile is very close to the initial profile of the parent stars. For two turnoff stars, this means that the core of the merger remnant is still mostly helium and that the object may not be able to remain on the main sequence for a very long time, since it is born with very little hydrogen left to burn at the center. In the case of collisions between two stars of very different masses, the chemical composition profiles of the merger remnants can be rather peculiar. For example, it often happens that the maximum helium abundance does not occur at the center of the remnant. These results are illustrated in Figure 1, which compares interior profiles of the final configurations following collisions between two stars with a mass ratio of 1/2 and between two identical stars, at various impact parameters.

At a qualitative level, these results can be understood very simply in terms of the requirement of convective stability of the final configurations. If entropy production in shocks could be neglected entirely (which may not be too unreasonable for the low-velocity collisions occurring in globular clusters), then one could predict the final composition profile simply by observing the composition and entropy profiles of the parent stars. Convective (dynamical) stability requires that the specific entropy $s$ increase monotonically from the center to the surface ($ds/dr > 0$, cf. Fig. 1) in the final hydrostatic equilibrium configuration. Therefore, in the absence of shock-heating, fluid elements conserve their entropy and the final composition profile of a merger remnant could be predicted simply by combining mass shells in order of increasing entropy, from the center to the outside. Many features of the results follow directly.

For example, in the case of a collision between two identical turnoff stars, it is obvious why the composition profile of the merger remnant remains very similar to that of the parent stars, since shock-heating is significant only in the outer layers of the stars, which contain a very small fraction of the total mass. The low-entropy, helium-rich material is concentrated in the deep interior of the parent stars, where shock-heating is negligible, and therefore it remains concentrated in the deep interior of the final configuration. For two stars of very different masses, the much lower-entropy material of the lower-mass star tends to concentrate in the core of the final configuration, leading to the unusual composition and temperature profiles seen in Figure 1(a). In essence, the smaller-mass star simply sinks in and settles at the center of the merger, while the higher-entropy, helium-rich material has been pushed out.

Sills, Bailyn & Demarque (1995) were the first to explore the consequences of blue stragglers being born unmixed. Using detailed stellar structure calculations, they compared the predicted colors ($U - B$ and $B - V$) of initially unmixed blue stragglers with observations. They concluded that some blue stragglers have observed colors that *cannot be explained* using unmixed initial models. Initially homogeneous models, however, can reproduce all the observations. In addition, unmixed models have much shorter main-sequence lifetimes than homogeneous models (because the cores of unmixed models have a very limited supply of hydrogen to burn), and there-fore may be incompatible with the observed *numbers* of blue stragglers. Thus we now have some indications from the observations that *significant mixing must take place* during the blue straggler formation process, and this is in apparent conflict with the latest results of hydrodynamic calculations. In addition, spectroscopic measurements of surface rotation rates of blue stragglers indicate that they are slow rotators (Mathys 1987, 1991), also in contradiction with the predictions of dynamical merger calculations. These problems are not peculiar to stellar collisions. Binary coalescence on a dynamical timescale also produces rapidly rotating merger remnants, and since the binary coalescence process is less dissipative than direct collisions (i.e., it creates less shock-heating), we expect even less hydrodynamic mixing in this case (Rasio & Shapiro 1995).

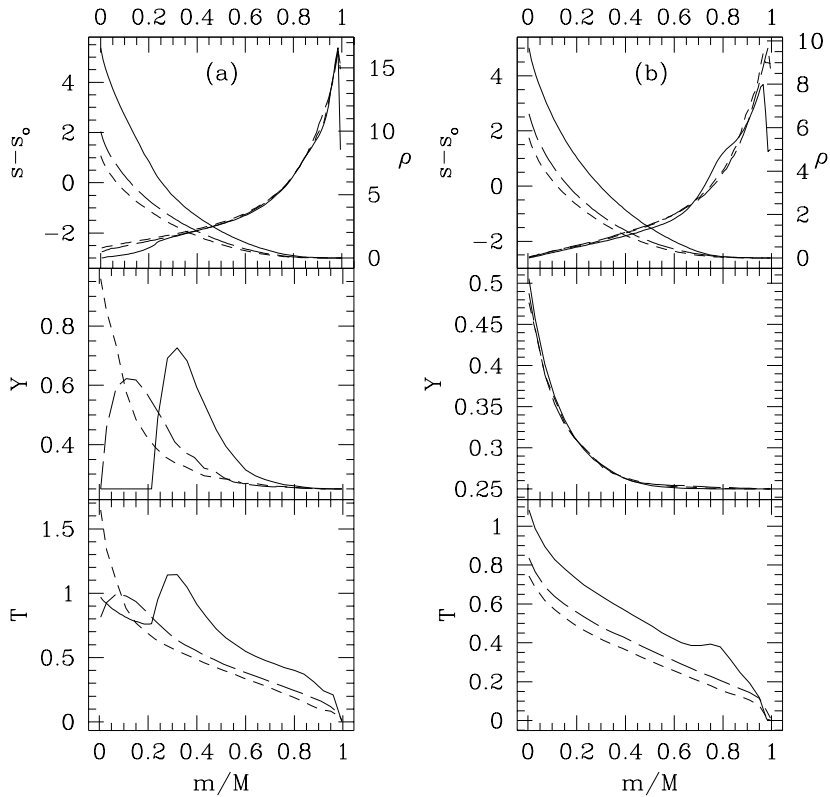

*Figure 1.* Final interior profiles for several merger remnants. The density $\rho$, specific entropy $s$ (up to a constant $s_o$), fractional helium abundance $Y$, and temperature $T$ are shown as a function of interior mass fraction. The masses of the two colliding stars were $M_1 = 0.8\,M_\odot$ and $M_2 = 0.4\,M_\odot$ in (a), and $M_1 = M_2 = 0.6\,M_\odot$ in (b). The solid, long-dashed and short-dashed curves correspond to initial trajectories with increasing periastron separations, $r_p/(R_1 + R_2) = 0$, 0.25, and 0.5, respectively ($R_1$ and $R_2$ are the initial stellar radii). The units are defined by $G = M_{TO} = R_{TO} = 1$, where $M_{TO} \simeq 0.8\,M_\odot$ and $R_{TO} \simeq 1\,R_\odot$ (i.e., the mass and radius of a turnoff main-sequence star). The specific entropy $s$ increases monotonically from the center to the outside, as required for convective stability (except in the outermost few percent of the mass, which have not yet reached hydrostatic equilibrium). Note the peculiar composition and temperature profiles in (a).

To resolve these apparent conflicts, it may be necessary to take into account processes that have not yet been incorporated into the theoretical models. Hydrodynamic calculations are of course limited to following the evolution of mergers on a dynamical timescale ($t_{dyn}$ ∼ hours for main-sequence stars) but are not capable of following processes taking place on a thermal timescale ($t_{th} \sim 10^7$ yr). The final configurations obtained at the end of hydrodynamic calculations (such as the ones illustrated in Fig. 1) are very close to hydrostatic equilibrium, but are generally *far from thermal equilibrium*. This is evident simply from the typical size of the merger remnants: the 95% mass radius at the end of the dynamical phase is typically several times the radius of a main-sequence star of the same total mass. Thus the object will need to contract (on its Kelvin time) before it can become a main-sequence star. In addition, the interior profiles of the merger remnants show clear evidence of departures from thermal equilibrium. A temperature gradient inversion (as seen in Fig. 1a) is a particularly clear sign. In addition, in the case of significantly off-axis collisions, it is found that the rapidly rotating final configurations are not barotropic (i.e.. the angular velocity is not simply constant on cylinders centered on the rotation axis), and therefore they must in general be out of thermal equilibrium (see, e.g., Tassoul 1978).

As the merger remnant contracts to the main sequence and evolves towards thermal equilibrium, many processes can lead to additional mixing of the fluid. These include convection (which is well known to occur during the evolution of ordinary pre-main-sequence stars), and, for rapidly rotating configurations, meridional circulation. These processes can lead not only to mixing, but also to loss of angular momentum and rapid spin-down through magnetic breaking (Leonard & Livio 1995). Even in regions where the convective (dynamical) stability criterion $ds/dr > 0$ is satisfied, *local thermal instabilities* (i.e., secular instabilities) can still occur. The small vertical oscillations (at the local Brunt-Väisälä frequency $\Omega_{BV} \propto [ds/dr]^{1/2}$) of a fluid element in such a region have amplitudes that grow unstably on a timescale comparable to the local radiative damping time (see, e.g., Kippenhahn & Weigert 1990). In a thermally unstable region, mixing will occur on this timescale. For example, in a region where there is a positive molecular weight gradient ($d\mu/dr > 0$) stabilized by a positive temperature gradient ($dT/dr > 0$), as seen in Figure 1 (solid lines in Fig. 1a), fingers of helium-rich material will tend to develop and penetrate the hydrogen-rich material below. Detailed calculations of the thermal relaxation phase ("pre-main-sequence blue straggler" evolution) incorporating a treatment of all relevant mixing processes will be necessary in order to develop a complete theoretical understanding of blue straggler formation.

At the same time, more extensive comparisons with the latest observational data will be necessary. The first study completed by Sills, Bailyn & Demarque (1995) focused on a small number of bright blue stragglers in the core of one particular cluster, NGC 6397. Preliminary results from another study, by Ouellette & Pritchet (1996), looking at a much larger sample of blue stragglers in the globular cluster M3, apparently lead to rather different conclusions. For the blue stragglers observed by HST in the core of M3 (Guhathakurta et al. 1994), this study finds better agreement with *unmixed* initial models. This is in direct contradiction with the results of Sills et al. (1995) if one assumes that blue stragglers in the cores of dense clusters are all formed by the same mechanism. On the other hand, Ouellette & Pritchet (1996) also find that the distribution of blue stragglers in the low-density outer region of M3 (Ferraro et al. 1993) agrees better with the predictions of *fully mixed* models. Clearly the final word on this question has not been said, but with the rapidly increasing quantity and quality of observational data on blue stragglers, and with several groups now working on improving the theoretical calculations and making detailed comparisons with observations, we can look forward to rapid progress in this area over the next few years.

**Acknowledgements**

I am very grateful to Piet Hut, Eiichiro Kokubo, Junichiro Makino, and Daiichiro Sugimoto for their hospitality in Tokyo. I thank Sverre Aarseth, Rosemary Mardling, and Frank Verbunt for many stimulating exchanges. I also thank Jamie Lombardi for his help in preparing the manuscript. Many of the results presented here were obtained from computations performed at the Cornell Theory Center, which receives major funding from the NSF and IBM Corporation, with additional support from the New York State Science and Technology Foundation and members of the Corporate Research Institute.